\documentclass[twocolumn]{aastex631}

\newcommand{\jme}[1]{\textcolor{orange}{#1}}

\usepackage{comment}
\usepackage{soul}
\usepackage{gensymb}
\usepackage{amsmath}
\usepackage{savesym}
\savesymbol{tablenum}
\usepackage{siunitx}
\usepackage{orcidlink}
\restoresymbol{SIX}{tablenum}
\usepackage{siunitx}

\graphicspath{{./}{figures/}}

\begin{document}

\title{A New Sample of Dwarf Galaxies with X-ray-Selected AGN Candidates from the eROSITA All-Sky Survey} 

\author[0009-0000-4535-6340]{John-Michael Eberhard \orcidlink{0009-0000-4535-6340}}
\affiliation{eXtreme Gravity Institute, Department of Physics, Montana State University, Bozeman, MT 59717, USA}
\author[0000-0001-7158-614X]{Amy E. Reines \orcidlink{0000-0001-7158-614X}}
\affiliation{eXtreme Gravity Institute, Department of Physics, Montana State University, Bozeman, MT 59717, USA}

\begin{abstract}
{
To investigate the population of massive black holes in dwarf galaxies, we conduct a systematic search for active galactic nuclei (AGNs) using data from the {first data release of the} eROSITA All-Sky Survey ({eRASS1}). We crossmatch dwarf galaxy positions in the NASA-Sloan Atlas with X-ray sources from eRASS and apply rigorous criteria to remove contaminants such as background sources, galaxies with dubious stellar masses, and X-ray binaries. {Potential contamination from ultraluminous X-ray sources is also assessed.} We ultimately assemble a sample of 27 X-ray-selected AGN {candidates} in dwarf galaxies with stellar masses $M_{*} = 10^{7.5-9.5} \; M_{\odot}$ and redshifts $z<0.15$.
An analysis of the host galaxy properties reveals that the AGN {candidates} are predominantly situated in systems with $g-r$ colors, star formation rates, and concentrations typical of the broader dwarf galaxy population. Detailed examination of the X-ray sources indicates that most AGN {candidates} in our sample are consistent with being located in the nuclear regions of their host galaxies and exhibit Eddington ratios ranging between $\lambda_{Edd} \sim 10^{-3}-10^{-1}$, with a few radiating at/above their Eddington limit. 
Notably, our methodology identifies 15 previously unreported AGNs {candidates},
highlighting the complementary nature of our approach {to the existing literature}. 
}
\end{abstract}

\section{Introduction} \label{sec:intro}
{Super}massive black holes (BHs) with masses $M_{BH}$$\sim 10^{6-9} M_\odot$ reside at the center of all large galaxies.
\citep[e.g.,][]{magorrian1998}. 
The formation pathways of {super}massive BHs remain a a subject of ongoing investigation, with several theoretical models proposed to explain their origin.
Proposed formation mechanisms include the collapse of Population III stars to form BH seeds (e.g., \citealt{bromm2003}), the 
growth of BHs through the merger of stars and smaller BHs within dense nuclear clusters 
(e.g., \citealt{port2004}; \citealt{gurkan2004}; \citealt{natarajan2021}), and the direct collapse of gas clouds in environments where fragmentation and star formation are suppressed (e.g., \citealt{lodato2006}; \citealt{begelman2006}). The relative contribution of these pathways to the cosmic black hole population remains uncertain, necessitating further observational constraints and theoretical refinement.


One  approach to differentiate between the different models is to study massive BHs 
in dwarf galaxies. Due to their relatively quiescent evolutionary histories, local dwarf galaxies are sensitive to the initial conditions of BH formation \citep[e.g.,][]{volonteri2008}, 
{which} makes them valuable for investigating the origins of \text{super}massive BHs.
In this study, we focus on identifying massive BHs in dwarf galaxies by searching for X-ray active galactic nuclei (AGNs).

X-ray observations are a powerful tool for detecting AGNs, particularly those with low Eddington ratios  (e.g., \citealt{gallo2008}; \citealt{hickox2009}) and/or those embedded in galaxies with significant star formation 
\citep[e.g.,][]{reines2014,reines2016,kimbro2021}. 
However, previous X-ray surveys have faced limitations due to constraints in sensitivity, angular resolution, and sky coverage. 
For instance, the  Roentgensatellit (ROSAT) all-sky survey, while groundbreaking in its time, achieved a modest average angular resolution of 30\arcsec 
\citep{trumper1984}. 

Despite these challenges, X-ray emission
has been successfully employed to identify AGNs in dwarf galaxies. 
For example, \cite{lemons2015} found 19 local dwarf galaxies with candidate AGNs 
using the Chandra Source Catalog, and \cite{birchall2020} found 61 candidates using the 3XMM catalog. 
 More recently, \cite{latimer2021} utilized the extended ROentgen Survey with an Imaging Telescope Array (eROSITA) Final Equatorial Depth Survey (eFEDS) to identify 6 AGN candidates in dwarf galaxies, with follow-up Chandra observations by \cite{sanchez2024} confirming X-ray emission consistent with AGNs in 2 of the candidates. 
 Importantly, \cite{latimer2021} also estimated that an all-sky survey by eROSITA could uncover as many as $\sim$1,300 AGNs in dwarf galaxies at low redshifts. 

The launch of the Spektrum Roentgen Gamma (SRG) observatory \citep{sunyaev2021}, equipped with the eROSITA X-ray telescope \citep{Predehl2021} represented a transformative step forward in X-ray observations. eROSITA provides the most sensitive X-ray all-sky survey to date, with sensitivity $\sim$25 times greater than ROSAT
in the $0.2–2.3$ keV band \citep{Predehl2021}. The data release of the eROSITA All-Sky Survey (eRASS) offers an extraordinary opportunity to search for X-ray AGNs in dwarf galaxies on a much larger scale.

{Several} recent studies have leveraged eRASS data to search for AGNs in dwarf galaxies. \cite{bykov2024} and \cite{sacchi2024} conducted analyses of the eastern galactic and western galactic hemispheres of eRASS, respectively, each identifying $\sim$80 AGN candidates in each region. {\cite{sacchi2024} utilized the first data release (eRASS1, \citealt{merloni2024}) while \cite{bykov2024} utilized the first four all-sky scans (eRASS:4). \cite{hoyer2024} also utilized eRASS:4 and searched for BHs located in galaxies' nuclear star clusters. They found 7 AGN candidates, of which 2 were located in galaxies with stellar masses $M_{*} < 10^{10} M_{\odot}$. \cite{burke2025} combined results from eRASS1, the Chandra Source Catalog \citep[CSC,][]{evans2024}, the XMM-Newton Serendipitous Source Catalog \citep{webb2020}, and the ROSAT all-sky survey point-source catalog \citep{boller2016} to make constraints on the  local ($<$50 Mpc) BH occupation fraction.} 

{In this work, we perform a complementary search, focusing on the western hemisphere of eRASS1. We perform a search analogous to that of \cite{sacchi2024}, but with a new set of selection criteria. We also utilize a different parent sample of galaxies that extend to higher redshifts, resulting in the identification of AGN candidates not reported in previous studies.}  

The paper is organized as follows: in \S2, we discuss our parent samples of X-ray sources and dwarf galaxies. In \S3, we detail the crossmatching procedure and the cleaning criteria  used to construct the final AGN sample. In \S4 we present an analysis of the properties of the AGNs and their host galaxies. We summarize our findings in \S5.

\section{Data}
\label{sec:Data}

\subsection{eROSITA/{eRASS1}}
\label{subsec:ERASS}
eROSITA is a wide-field X-ray telescope designed to operate in the $0.2–10$ keV 
energy range, with an average angular resolution of $\sim$15\arcsec\ at 1.5 keV and a nominal positional accuracy of 3\arcsec\ \citep{Predehl2021}. The sensitivity of eROSITA varies depending on the location in the sky; however, \citet{merloni2024} report that the first full-sky scan (eRASS1) has a median flux limit (at 50\% completeness) of  $F_{\mathrm{0.5-2 \; keV}} > 5\times10^{-14}\;\mathrm{erg\;s^{-1}\;cm^{-2}}$.

Beginning in 2019, eROSITA completed $\approx4.4$ all-sky X-ray scans as part of the all-sky survey eRASS. Observations were suspended in 2022 due to geopolitical tensions and the data are shared equally between German and Russian scientists, with the western galactic hemisphere ($180 < l < 360\degree$) being managed by the German eROSITA consortium \citep{merloni2024}. 
For this work, we utilize the main catalog of eRASS of the western hemisphere, which is publicly available at this time.
This catalog contains all point-like and extended X-ray sources with detections 
in the $0.2-2.3$ keV band, 
with a total of $\sim$930,000 sources.

\subsection{NASA-Sloan Atlas (NSA)}
\label{subsec:NSA}
We utilize the NASA-Sloan Atlas (NSA, v1\textunderscore0\textunderscore1) to establish a parent sample of dwarf galaxies. The latest version of the NSA is based on observations from SDSS DR11, with a sky coverage area of 14,555 square degrees that extends to a redshift of $z < 0.15$. The catalog lists galaxies and their $ugriz$ magnitudes, which are derived via images from the Sloan Digital Sky Survey (SDSS) and the Galaxy Evolution Explorer (GALEX, \citealt{york2000}; \citealt{blanton2005}). The stellar masses ($M_{*}$) and the halflight radii ($r_{50}$) of the galaxies are also included in the catalog, derived from elliptical Petrosian photometry. The masses are calculated using the \textit{kcorrect} package \citep{blanton2007} and are given in units of $M_\odot h^{-2}$. We employ $h$ = 0.73 throughout the paper to calculate the stellar masses of the galaxies.

\section{Sample Selection}
\label{sec:sample_selection}

\subsection{Dwarf Galaxy Sample}
\label{subsec:dwarf_gal_sample}
To assemble our parent sample of dwarf galaxies, we 
restrict the NSA to galaxies with stellar masses $M_{*}<3\times 10^9 \; M_\odot$. This criterion results in a sample of 63,656 dwarf galaxies. 
{Of the 63,656 dwarf galaxies identified in the NSA, 29,030 (45.6\%) are situated within the western galactic hemisphere, coinciding with the eRASS1 survey footprint. We compare this sample to that utilized in \cite{sacchi2024}, which based its parent sample on the HECATE catalog: an all-sky catalog wherein approximately half of the entries fall within the eRASS1 footprint. Despite HECATE’s broader sky coverage, it contains a comparatively lower number of dwarf galaxies, with only 5,775 objects exhibiting masses consistent with this classification. Consequently, our catalog identifies a significantly larger population of dwarf galaxies than HECATE, highlighting the potential for discovering previously unrecognized candidates.}

We crossmatch {our sample of dwarf galaxies} to the X-ray sources in {eRASS1}, selecting each galaxy where the separation between the galaxy's optical center and the X-ray source is less than twice the galaxy's half-light radius (2$r_{50}$).  To minimize the inclusion of galaxies with inaccurately large $r_{50}$ values, we further constrain the matches to galaxy-source pairs with crossmatching radii smaller than 15\arcsec, corresponding to the angular resolution of eROSITA and the crossmatching radius employed by \cite{sacchi2024}. Applying these criteria results in a sample of 68 galaxies with X-ray sources located within 2$r_{50}$ and 15$\arcsec$ of their optical centers.

\subsection{Removal of Background AGNs}
\label{subsec:background}
\begin{figure}
\includegraphics[width=\columnwidth]{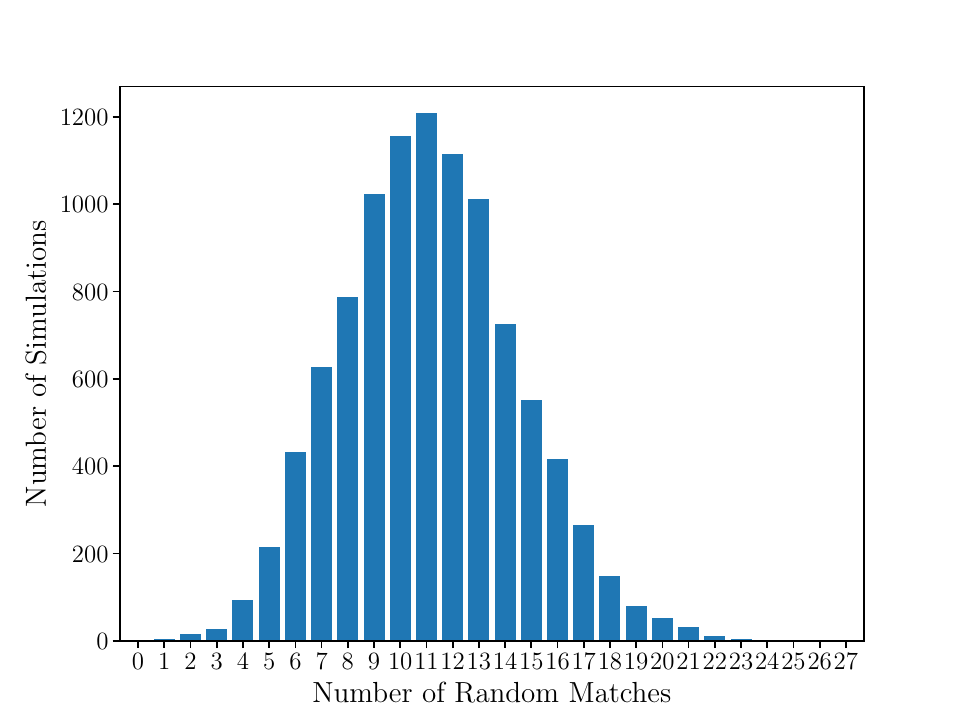}
 \caption{Histogram of the number of matches found for 10,000 simulations that matched randomly generated sky coordinates to the locations of the X-ray sources from {eRASS1}.}
\label{fig:coinc_histo}
\end{figure}

We consider the likelihood of background X-ray sources matching with dwarf galaxy coordinates by coincidence. We perform a Monte Carlo simulation where we randomly generate 63,656 coordinate pairs within the NSA footprint to represent randomly placed galaxies. We match these coordinates to the actual X-ray source locations, using the 2$r_{50}$ values from the original dwarf galaxy sample as crossmatching radii. We then restrict the sample to only include galaxies with crossmatching distances below 15\arcsec\, as was done in our selection process. We repeat the procedure 10,000 times.
The resulting distribution of coincidental matches, as seen in Figure \ref{fig:coinc_histo}, has a mean of 11.1 
and a standard deviation of 3.3, yielding a 95\% confidence interval of $4.5-17.7$ matches. We therefore expect between 4 and 18 of the X-ray sources in our sample of 68 dwarf galaxies to be coincidental matches.

To find and remove the background sources in our sample, we crossmatch the X-ray sources in {eRASS1} with the MILLIQUAS catalog \citep{flesch2023}, as was done in \cite{sacchi2024}. The MILLIQUAS catalog contains nearly one million quasars, along with additional galaxies that have been identified as AGNs. We find that 7 X-ray sources
are included in the MILLIQUAS catalog with redshifts much larger than the redshifts reported in the NSA. These sources are removed from our sample for being background quasars. We also match our X-ray sources to the SDSS DR16 list of quasars \citep{lyke2020}, as was done in \cite{bykov2024}, but find no additional quasars. 

We also search for background sources by performing a visual inspection of the X-ray source positions relative to their host galaxies using images from the Dark Energy Camera Legacy Survey (DECaLS; \citealt{dey2019}). We assess whether any X-ray sources exhibit significant spatial offsets from their associated galaxies, defining a source as offset if its X-ray position, including its positional uncertainty (as reported in the $RA\_DEC\_ERR$ column of the {eRASS1} catalog) does not overlap at all with the  galaxy it crossmatched with.  While our initial crossmatching procedure incorporated the half-light radii to mitigate such misassociations, the asymmetric morphology of many dwarf galaxies introduces additional complexity, allowing some sources located within $2r_{50}$ of the galaxy center to still be positioned well outside {the galaxy itself}. 
We find 6 sources that are offset from their host galaxy and eliminate them from our sample. 
During the inspection, we also remove one X-ray source which has a clear optical point-like counterpart, consistent with a background quasar. 



In total, we identify and exclude 14 X-ray sources as likely background contaminants. This result is consistent with the expected range of $4–18$ coincidental matches predicted by our Monte Carlo simulations.


\subsection{Unreliable Stellar Masses}
\label{subsec:mass_gals}
To ensure our sample includes only dwarf galaxies, we search for potential inaccuracies in the stellar masses reported by the NSA, which are dependent on redshifts. If the redshifts in the NSA are incorrect, the derived stellar masses will also be erroneous. Therefore, to remove all galaxies with incorrect stellar masses, we include only galaxies with verifiable redshifts in our final AGN sample.

For each galaxy, we examine its spectral data available in the SDSS catalog and the NASA Extragalactic Database (NED). We confirm the reported redshifts by ensuring that the positions of prominent emission lines in the spectra correspond to the redshifts provided in the NSA. Sixteen galaxies are excluded because their redshifts cannot be verified. This is due to the absence of spectra in the SDSS or NED, the inclusion of spectra for objects other than the galaxy (e.g., nearby stars), or unreliable spectral fits. Without confirmation that these galaxies have stellar masses consistent with dwarf galaxies, we remove them from our sample.

Additionally, we remove 3 nearby galaxies 
known to have masses exceeding the dwarf galaxy threshold. These galaxies likely had their masses underestimated in the NSA due to photometric fragmentation, where galaxies of a large size are measured by different fibers, and one of the fibers provides a mass within the dwarf regime. This issue was also observed in the dwarf galaxy sample studied by \cite{bykov2024}. 

We remove a total of 19 galaxies from our sample for having unreliable mass estimates. We consider the remaining 35 galaxies to be bona fide dwarf galaxies.

We also identify one galaxy that, despite having a reliable redshift measurement,   has an unrealistically low stellar mass of $2.3\times 10^{6} \; M_\odot$. We address this by recalculating the stellar mass using the mass-to-light ratio in the $i$-band ($M/L_{i}$), derived from the $g-i$ color following \cite{zibetti2009}: 

\begin{equation}
\mathrm{log}\left(M/L_{i}\right) = 1.032\,(g-i) -0.963
\end{equation}

\begin{figure}
\includegraphics[width=\columnwidth]{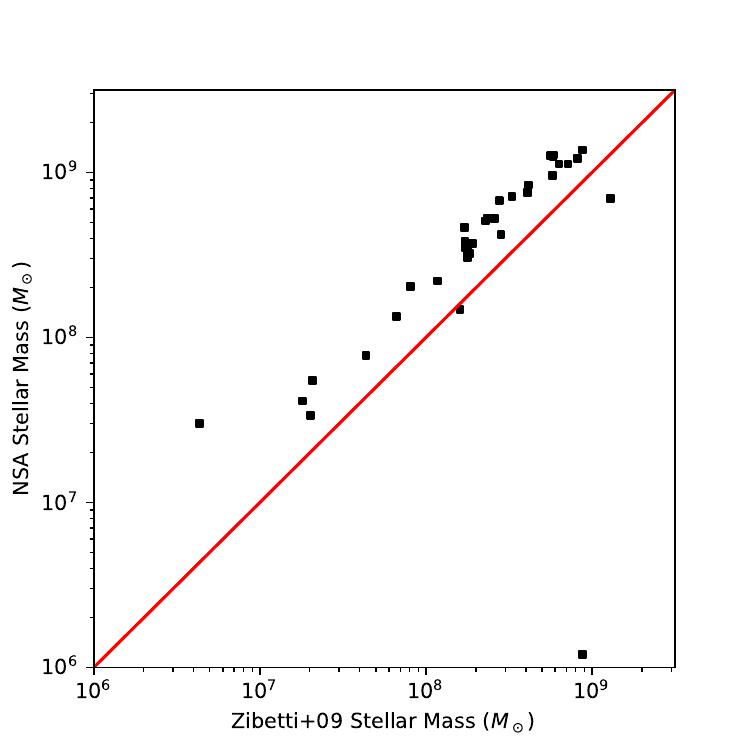}
 \caption{Comparison of stellar masses from the NSA to stellar masses estimated using mass-to-light ratios 
 as a function of $g - i$ color,
following \cite{zibetti2009}. The red line shows the one-to-one relation. 
} 
\label{fig:mass_compare}
\end{figure}

As noted by \cite{reines2015}, masses derived using this approach generally align with the NSA masses, with a median offset of only 0.06 dex. 
However, the NSA masses have larger scatter at the low-mass end of the relation, as seen in Figure 6 of \cite{reines2015}.
For our sample of dwarf galaxies, we observe a median offset of 0.28 dex (see Figure \ref{fig:mass_compare}), with the NSA masses typically being larger than the \cite{zibetti2009} masses for dwarf galaxies, {as was seen in \cite{reines2015}}.
However, for this particular galaxy, the mass derived from the \cite{zibetti2009} method is significantly higher, with a value of $1.6 \times 10^{9} M_\odot$. Since the recalculated mass is still within the dwarf galaxy regime, we adopt the higher value for subsequent analyses, as it is more consistent with expectations than the NSA-reported mass.

{We note that the determination of stellar masses using the \cite{zibetti2009} method is inherently dependent on the $i$-band luminosity, which itself is dependent on redshift. Therefore, the galaxies we excluded due to unreliable redshift measurements also lack reliable mass estimates under this method and remain excluded from our analysis.}

\subsection{Emission from X-ray Binaries}
\label{subsec:HMXBs_XRBs}
{Individual  high-luminosity X-ray binaries (XRBs) can produce compact, point-like X-ray emission that may mimic AGN signatures and have the potential to contaminate our sample. Following the approach of \cite{sacchi2024}, we apply the 0.2–2.3 keV X-ray luminosity threshold of $L_{X} < 10^{39} \; \mathrm{erg \; s^{-1}}$  to our sample to exclude such sources. Unlike in \cite{sacchi2024}, none of the objects in our sample fall below this luminosity threshold, and thus no sources are removed. Consequently, we expect minimal contamination from individual resolved XRBs in our sample.}

Unresolved populations of XRBs within host galaxies may also mimic the X-ray emission from AGNs, since the angular resolution of eROSITA is comparable to the physical extent of the galaxies in our sample. To assess and mitigate potential contamination from XRB populations, we compare the observed X-ray luminosities to predicted contributions from XRB populations using empirical scaling relations. 

The total X-ray luminosity from XRBs is given by the sum of contributions from low-mass XRBs (LMXBs) and high-mass XRBs (HMXBs):
\begin{equation}
\mathrm{L_{XRB}} = \mathrm{L_{LMXB}} + \mathrm{L_{HMXB}}
\end{equation}
Since LMXB luminosity scales linearly with stellar mass ($M_{*}$), and HMXB luminosity scales linearly with the star formation rate \citep[SFR; e.g.,][]{colbert2004}, this relation is often parameterized as:
\begin{equation}
\mathrm{L_{XRB}} = \alpha \; M_{*} + \beta \; \mathrm{SFR}
\end{equation}
where $\alpha$ and $\beta$ are empirically derived coefficients. The relation exhibits a scatter of 0.34 dex for local galaxies \citep{lehmer2010}. 

The best-fit coefficients $\alpha$ and $\beta$ exhibit variation across different galaxy samples and depend on the method used to derive SFR. To obtain a robust estimate of XRB contamination and ensure a conservative AGN selection, we predict the XRB luminosities using $\alpha$ and $\beta$ values from three independent studies.

First, we adopt the $\alpha$ and $\beta$ values from \cite{lehmer2010}.
As in \cite{lehmer2019} and \cite{sacchi2024}, we calculate SFR using the following equation from \cite{kennicutt2012}:
 \begin{equation}
 \label{eq:sfr_ir_fuv}
     \mathrm{log \; SFR}(M_\odot \; \mathrm{yr}^{-1}) = \mathrm{log \; L(FUV)_{corr}} - 43.35
 \end{equation} 
Here, $\mathrm{L(FUV)_{corr}}$ is the corrected far-ultraviolet (FUV) luminosity, which is calculated by combining the observed FUV and 25$\mu$m luminosities, using the following relation from \cite{hao2011}:
 \begin{equation}
  \label{eq:fuv_corr}
    \mathrm{\mathrm{L(FUV)_{corr}} = L(FUV)_{obs} + 3.89 \; L(25\mu m)}
 \end{equation}
We use the GALEX $k$-corrected FUV magnitudes provided in the NSA to calculate $\mathrm{L}(\mathrm{FUV})_{\mathrm{obs}}$.
Given that the ratio between the 25 $\mu$m and 22 $\mu$m flux densities is around order one \citep{jarrett2013}, we use the 22 $\mu$m (W4 band) magnitudes from the Wide-field Infrared Survey Explorer \citep[WISE;][]{allwisecat}\footnote{This publication makes use of data products from the Wide-field Infrared Survey Explorer, which is a joint project of the University of California, Los Angeles, and the Jet Propulsion Laboratory/California Institute of Technology, funded by the National Aeronautics and Space Administration.} to calculate $\mathrm{L(25\mu m)}$. 

Since \cite{lehmer2010} derived the $\mathrm{L_{XRB}}$ relation for the $2–10$ keV energy band, we convert the observed $0.2–2.3$ keV luminosities from {eRASS1} to $2–10$ keV using a correction factor of -0.06 dex (a factor of $\sim$0.87). We derive this value using the Chandra-based Portable, Interactive Multi-Mission Simulator (PIMMS) \footnote{https://cxc.harvard.edu/toolkit/pimms.jsp} assuming a photon index of $\Gamma = 2$ and a galactic absorption of $N_{H}=3\times10^{20} \; \mathrm{cm}^{-2}$. 

The comparison of the observed $2–10$ keV luminosities to the predicted luminosities from XRBs using the \cite{lehmer2010} scaling relation is shown in the left panel of Figure \ref{fig:xrbs}. We define any objects with observed X-ray luminosities that fall within 1$\sigma$ scatter of the $\mathrm{L_{XRB}}$ relation to be consistent with XRBs.
Under this criterion, only 1 galaxy is classified as having emission consistent with XRBs using the \cite{lehmer2010} relation.

We repeat the procedure using the $\alpha$ and $\beta$ values derived from the {full} sample of galaxies from \cite{lehmer2019}. 
Using the \cite{lehmer2019} relation, 7 galaxies are shown to have emission consistent with XRB populations, one of which is the same galaxy that was identified using the \cite{lehmer2010} relation (see the middle panel of Figure \ref{fig:xrbs}).

{We lastly apply the analysis of \citet{geda2024}. Although explicit values of the scaling coefficients $\alpha$ and $\beta$ are not quoted, we calculate them from the luminosity functions via 
\begin{equation}
    \beta = \frac{1}{\mathrm{SFR}}\int_{L_{lo}}^{L_c} \frac{dN_{HMXB}}{dL} \; L \; dL 
\end{equation}
and
\begin{equation}
    \alpha = \frac{1}{\mathrm{M_{*}}}\int_{L_{lo}}^{L_c} \frac{dN_{LMXB}}{dL} \; L \; dL 
\end{equation}
Here, $L_{lo}$ and $L_c$ denote the lower and cutoff luminosities, respectively, as reported for the full dwarf galaxy sample in \citet{geda2024}. The functional forms of the HMXB and LMXB XLFs, $\frac{dN_{\mathrm{HMXB}}}{dL}$ and $\frac{dN_{\mathrm{LMXB}}}{dL}$, are likewise taken from their analysis.}

\begin{figure}
\includegraphics[width=\columnwidth]{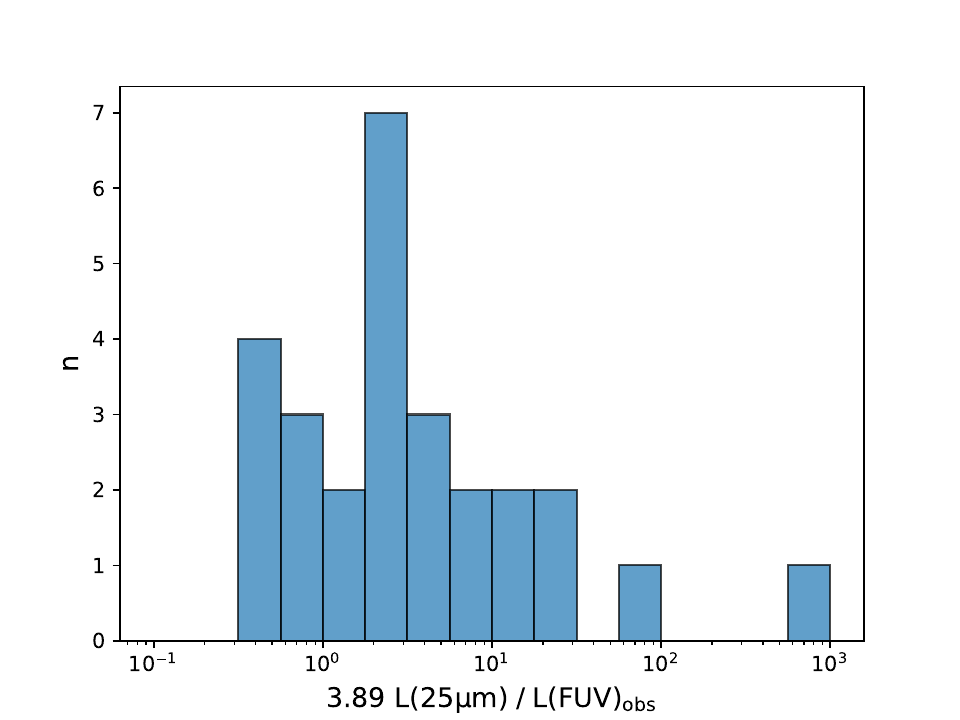}
\caption{{Ratio of the IR correction applied to the FUV luminosities (see Equation \ref{eq:fuv_corr}) relative to the observed FUV luminosities for galaxies in our final sample.}}
\label{fig:l25_lfuv}
\end{figure}

\begin{figure}
\includegraphics[width=\columnwidth]{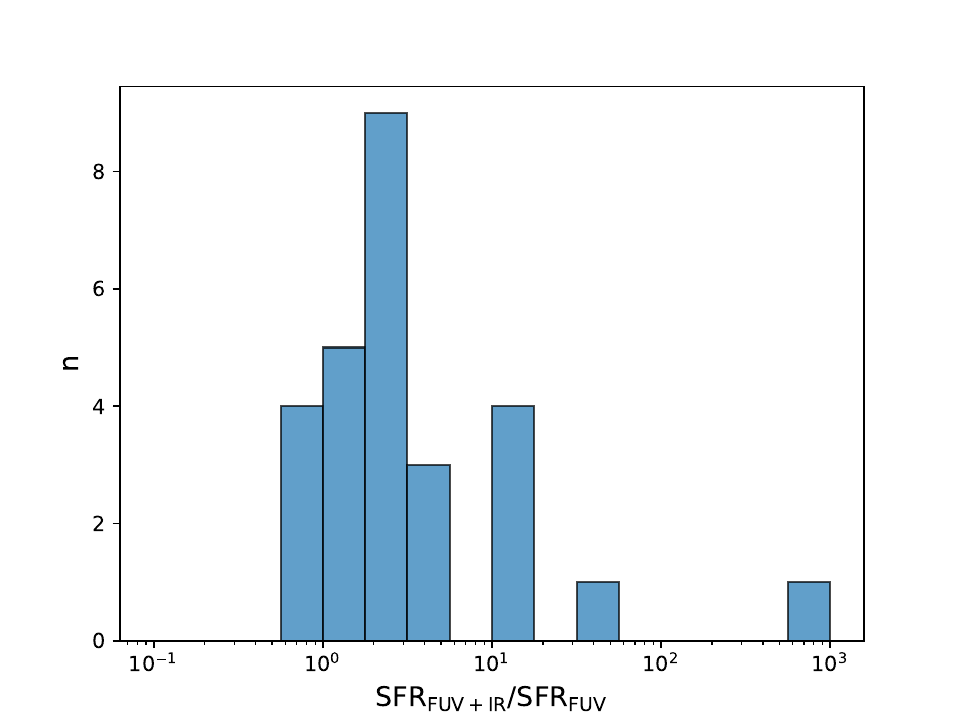}
\caption{{Ratio of SFRs derived from IR-corrected FUV luminosities (Equation \ref{eq:sfr_ir_fuv}) to those derived from FUV luminosities alone (Equation \ref{eq:sfr_fuv}) for the galaxies in our final sample.}}
\label{fig:sfr_ratio}
\end{figure}

\begin{figure*}
\centering
\includegraphics[width=\textwidth]{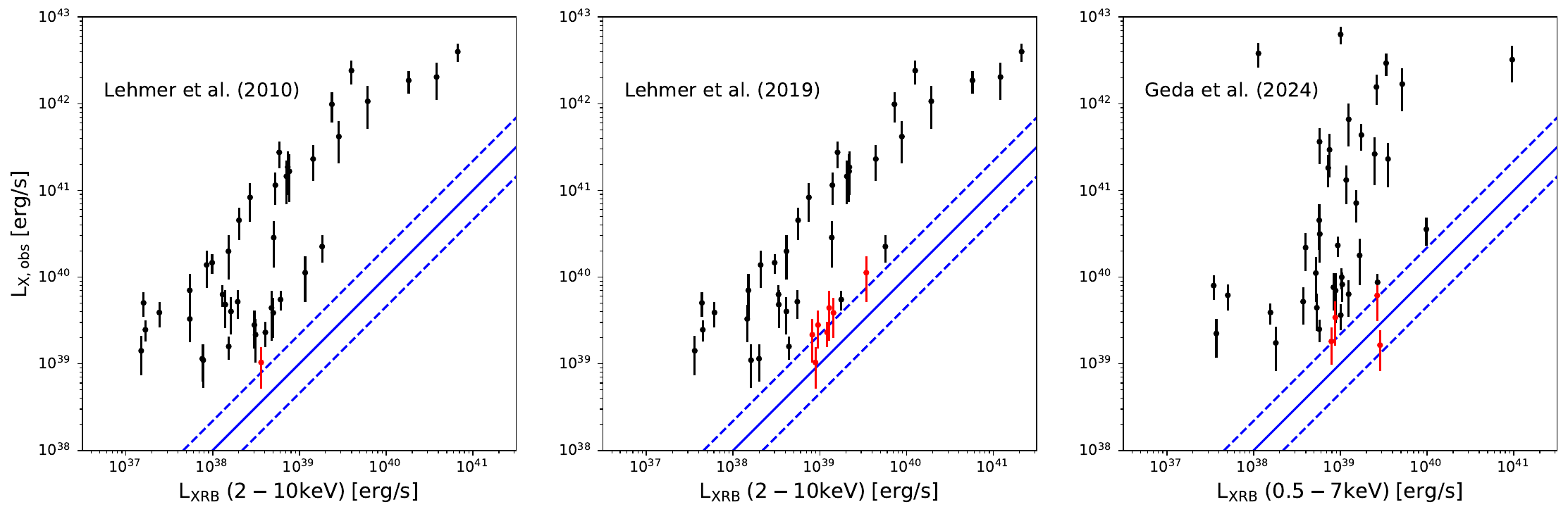}
\caption{Comparison of observed X-ray luminosities versus the expected luminosity from XRBs, using the XRB relation from 3 different papers. The blue solid line shows the one-to-one relation, with the blue dashed lines showing the scatter in the XRB relation. The 1$\sigma$ error bars of the actual X-ray luminosities are shown. Red points indicate galaxies that fall within 1$\sigma$ scatter of the XRB relation, indicating that they have emission consistent with XRB populations. Left: XRB relation from \cite{lehmer2010}. 
 Middle:  XRB relation from \cite{lehmer2019}. 
 Right:  XRB relation from \cite{geda2024}. 
 } 
\label{fig:xrbs}
\end{figure*}

{While their sample of dwarf galaxies is comparable to ours in stellar mass, \citet{geda2024} derived SFRs exclusively from FUV luminosities, adopting:}
\begin{equation}
\label{eq:sfr_fuv}
    \mathrm{SFR_{FUV}} = 1.4 \times 10^{-28} \, L(\mathrm{FUV})_{\nu},
\end{equation}
{where $L(\mathrm{FUV})_{\nu}$ is the FUV luminosity density ($\mathrm{erg \; s^{-1} \; Hz^{-1}}$). This calibration was applied without an infrared correction under the assumption of negligible dust attenuation in their sample. For our final sample of galaxies, however, the infrared correction (Equation \ref{eq:fuv_corr}) is frequently comparable to, or greater than, the observed FUV luminosity (Figure \ref{fig:l25_lfuv}), indicating that the dust attenuation is not negligible. As a result, the values of SFR$_{\mathrm{FUV}}$ can therefore differ significantly from the infrared-corrected SFRs for the galaxies in our final sample 
(Figure \ref{fig:sfr_ratio}). Given these differences, we complement the \citet{geda2024} relations with those of \citet{lehmer2010,lehmer2019} to ensure a more robust characterization of the XRB populations in our final sample.} 

We also note that the \cite{geda2024} scaling relations are derived for the $0.5–7$ keV band. To match this band, we convert the luminosities from {eRASS1} to  $0.5–7$ keV luminosities using a correction factor of 0.13 dex (a factor of $\sim$1.35), which we calculate using PIMMS. 

Using the scaling relation derived from \cite{geda2024}, 4 galaxies are found to have emission consistent with XRBs (right panel of Figure \ref{fig:xrbs}). Three of the galaxies were identified with the \cite{lehmer2010,lehmer2019} relations, and one is uniquely classified as having XRB-consistent emission by the \cite{geda2024} relation.

Utilizing three independent methods, we identify 8 galaxies whose X-ray luminosities are consistent with emission from XRB populations. As a result, these sources are excluded from our final AGN sample. However, we acknowledge that consistency with XRB emission does not definitively exclude the presence of an AGN.
As such, we provide images and relevant properties of these XRB-consistent sources in the Appendix for potential follow-up studies.

\subsection{Ultraluminous X-ray Sources}
As with any X-ray-selected sample, our dataset is susceptible to contamination  from ultraluminous X-ray sources (ULXs), defined as off-nuclear point sources exhibiting X-ray luminosities of $L_{X} > 10^{39} \; \mathrm{erg \; s^{-1}}$. 
Early interpretations suggested that ULXs may represent massive BH candidates (e.g., \citealt{colbert1999,liu2005}). However, subsequent studies have demonstrated that many ULXs are consistent with originating from the high luminosity tail of the HMXB population, particularly systems with stellar-mass BHs and neutron stars accreting at super-Eddington rates (e.g., \citealt{king2001,feng2011,bachetti2014,king2016,roberts2016,kayanikhoo2025}). 
{However, such ULXs are rarely observed with luminosities exceeding $L_X \sim 10^{41} \; \mathrm{erg \; s^{-1}}$ \citep[e.g.,][]{grimm2003,mineo2012,lehmer2019}}.

To mitigate contamination from ULXs,
we first search for known ULXs by crossmatching our sample with the Chandra and XMM-Newton archival databases\footnote{\url{https://cda.harvard.edu/chaser/mainEntry}, \url{https://nxsa.esac.esa.int/nxsa-web/\#search}}, as was done in \citet{sacchi2024}. This search yields six galaxies with archival detections, all of which were previously classified as massive BHs in the literature (see \S\ref{subsubsec:remaining_lit}). One galaxy was also identified as a ULX by \citet{thygesen2023}, based on its significant offset from the optical center. However, as discussed in \S\ref{subsec:offsets}, such an offset may also be indicative of a wandering BH. This source was also classified as a BH candidate by \citet{lemons2015} and \citet{sacchi2024}. Since all the matched sources were consistent with massive BHs, we keep them in our final candidate sample.

{We further assess the potential contamination from ULXs by estimating the expected number of ULXs ($N_{\mathrm{ULX}}$) arising from HMXBs per SFR, based on the HMXB X-ray luminosity function (XLF) presented in \cite{lehmer2019}. Specifically, we adopt the XLF fit to the full galaxy sample in that study. Although \cite{geda2024} also provided an HMXB XLF, their calibrations were based on SFRs derived under the assumption of negligible dust extinction. Since our sample includes galaxies where dust corrections are significant (Section \ref{subsec:HMXBs_XRBs}), the resultant $N_{\mathrm{ULX}}$/SFR values from \cite{geda2024} do not apply well to our sample and we do not include them in our analysis.}



{Assuming a minimum ULX luminosity of $L_X = 10^{39} \; \mathrm{erg \; s^{-1}}$ and integrating the \citet{lehmer2019} HMXB XLF above this value yields $N_{\mathrm{ULX}}/\mathrm{SFR} = 0.6 \; (M_\odot,\mathrm{yr}^{-1})^{-1}$. 
However, each galaxy has a limiting {\it detectable} X-ray luminosity given its distance, which must also be taken into account.
To estimate the number of {\it detectable} ULXs in our parent sample of dwarf galaxies in the western hemisphere of the NSA, we calculate, for each galaxy,
the minimum X-ray luminosity at which an X-ray source could be detected using the average eRASS flux limit from \citet{merloni2024} and the NSA redshift. We integrate the HMXB XLF for each galaxy, starting from its minimum detectable luminosity (or from $10^{39} \;\mathrm{erg \; s^{-1}}$ when the minimum detectable luminosity is lower than this) to obtain the value of $N_{\mathrm{ULX}}/\mathrm{SFR}$ for the galaxy. We then multiply by the galaxy's SFR to get $N_{\mathrm{ULX}}$. Summing over all the galaxies in the parent sample yields $N_{\mathrm{ULX}} = 20^{+3}_{-4}$.}

{If the estimate of $\sim 20$ ULXs is taken at face value, then approximately 30\% of the 68 eRASS sources that were crossmatched to NSA dwarf galaxies could plausibly be attributed to ULXs. We have applied several criteria to mitigate ULX contamination in our final sample, and the 8 galaxies excluded from our sample for having sources consistent with XRB-dominated emission represent the most likely ULX hosts. However, even with their removal, up to $\sim 12$ ULXs may be present among the remaining 60 crossmatched sources. 
In the extreme case, all $\sim 12$ could persist in the final sample, although the 9 with X-ray luminosities above $L_X > 10^{41} \; \mathrm{erg \; s^{-1}}$ are unlikely to be ULXs. We note that distinguishing ULXs from accreting massive BHs based solely on X-ray properties is inherently challenging and robust classification will ultimately require multi-wavelength follow-up. The sources presented here should therefore be regarded as candidates: possible AGNs that will require further observations to confirm their nature.}

\section{AGN Sample}
\label{sec:finalSample}
\begin{figure*}[t]
\centering
\includegraphics[height=22cm]{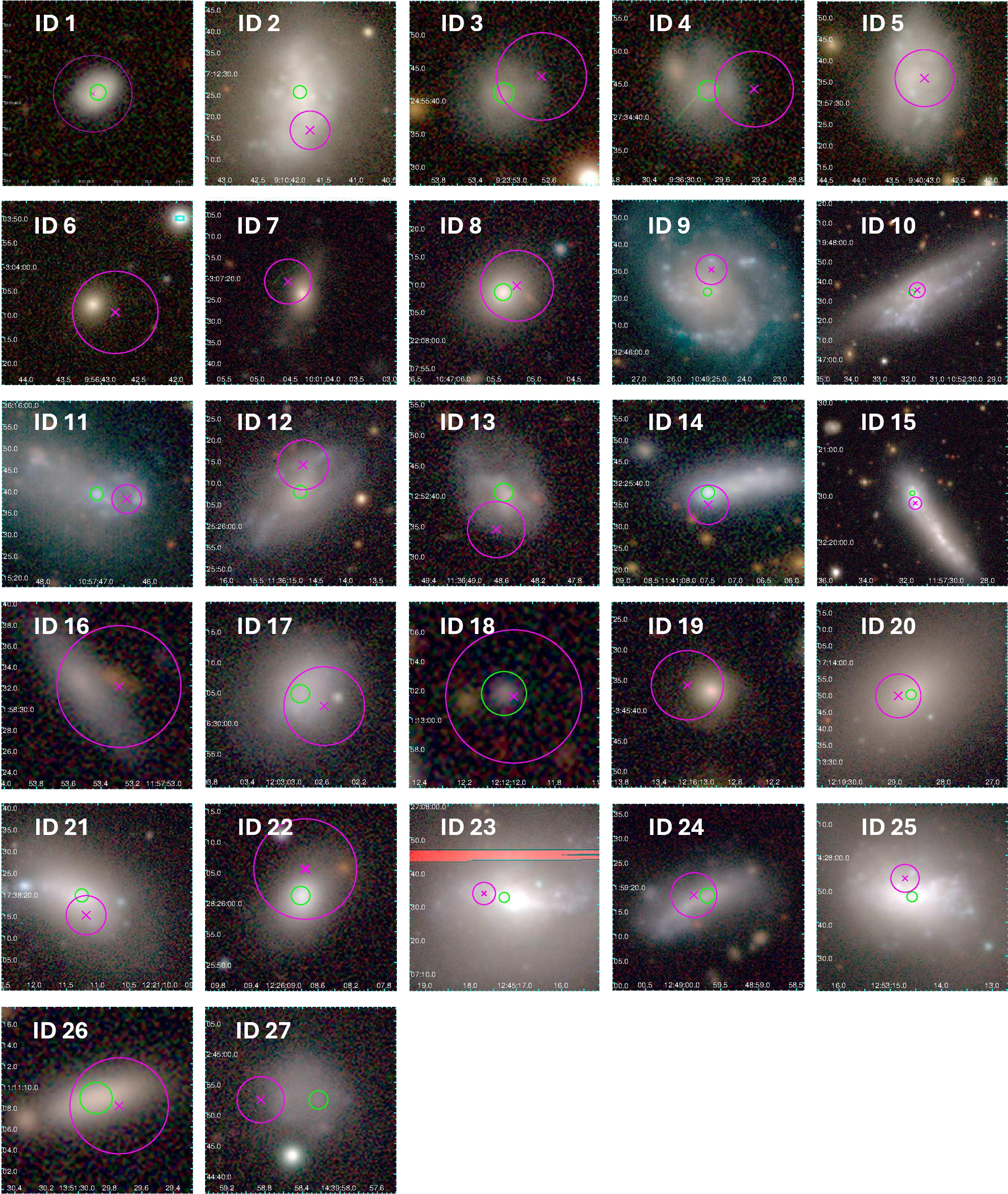}
 \caption{
 $grz$-band DECaLS images of the final sample of dwarf galaxies with X-ray AGNs. The positions of the X-ray sources from {eRASS1} are shown as magenta crosses, with magenta circles showing the positional uncertainties. The smaller green circles identify the locations of the SDSS spectral fibers (with 1.5\arcsec radii) for galaxies with corresponding SDSS spectra.}
\label{fig:decals_pics}
\end{figure*}

With the exclusion of background sources, galaxies with unreliable stellar masses, and potential XRB populations, a total of 41 galaxies are removed from our initial sample of 68 galaxies identified in the NSA and {eRASS1}. The remaining 27 constitute our final sample of dwarf galaxies that are candidates for X-ray AGN hosts. 
We provide DECaLS images of the 27 galaxies in our AGN sample in Figure \ref{fig:decals_pics} and a summary of their properties in Table \ref{tab:erass_agns}. For ease of reference, we assign each galaxy an identification number between 1 and 27.


\subsection{Host Galaxy Properties}
\label{subsec:host_gals}
\begin{figure}
\includegraphics[width=\columnwidth]{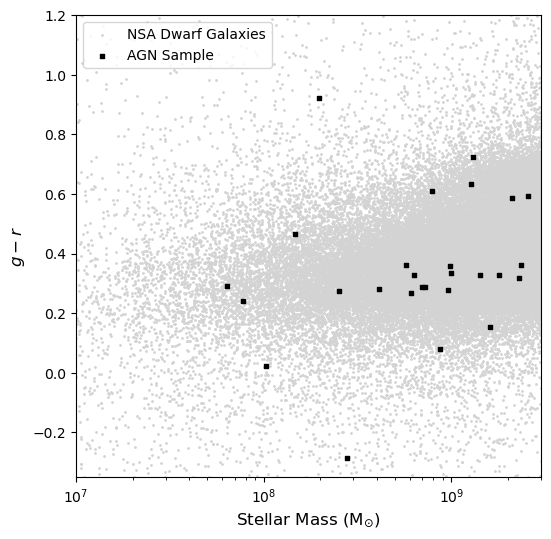}
 \caption{Stellar masses and $g-r$ colors of the host galaxies in our final sample of AGNs (black) and the full population of dwarf galaxies in the NSA (gray). The AGN hosts largely have similar masses and colors to the general population, with notable exceptions being IDs 7 and 18 which have notably redder and bluer $g-r$ colors than the full population, respectively.} 
\label{fig:mass_color}
\end{figure}

We assess the characteristics of the galaxies in our AGN sample by comparing their stellar masses and $g-r$ colors to the full population of dwarf galaxies in the NSA (Figure \ref{fig:mass_color}). The majority of the AGN hosts exhibit masses and colors consistent with the broader dwarf galaxy population. Notably, one galaxy (ID 7) displays a significantly redder color and one galaxy (ID 18) has a markedly bluer color. This distribution suggests that the X-ray method for identifying AGNs in dwarf galaxies is not strongly biased by host galaxy color.

\begin{figure}
\includegraphics[width=\columnwidth]{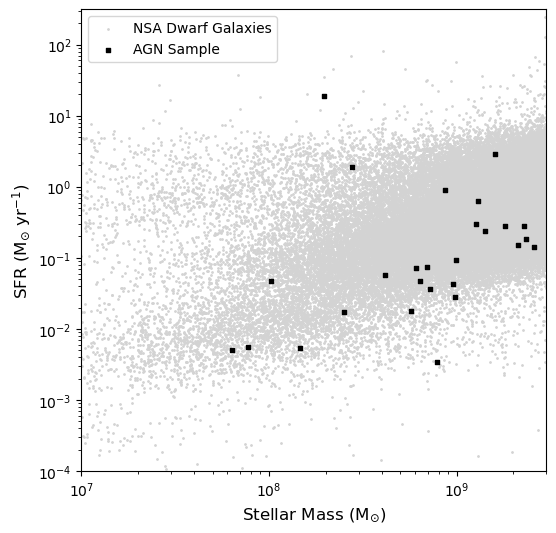}
 \caption{Stellar masses and SFRs of the host galaxies in our final sample of AGNs (black) and the full population of dwarf galaxies in the NSA (gray). One galaxy (ID 7) has an SFR that falls above the typical distribution of SFRs for dwarf galaxies. The remaining AGN hosts have similar SFRs to the general population with a slight bias towards lower SFRs, possibly the result of star formation being suppressed by AGNs.} 
\label{fig:sfr_mass}
\end{figure}

The SFRs of our X-ray AGN sample (calculated using Equation \ref{eq:sfr_ir_fuv}) are similarly compared to the to the full NSA dwarf galaxy population (Figure \ref{fig:sfr_mass}). 
The SFRs of the AGN host galaxies generally align with the broader dwarf galaxy population, though a subtle trend toward lower-than-average SFRs is observed for galaxies with stellar masses $M_{*} < 10^{9} \; M_\odot$.
This trend may indicate a potential role of AGN activity in suppressing star formation within lower-mass systems.
A notable exception is ID 7, which displays a significantly elevated SFR relative to the typical dwarf galaxy population.


\begin{figure}
\includegraphics[width=\columnwidth]{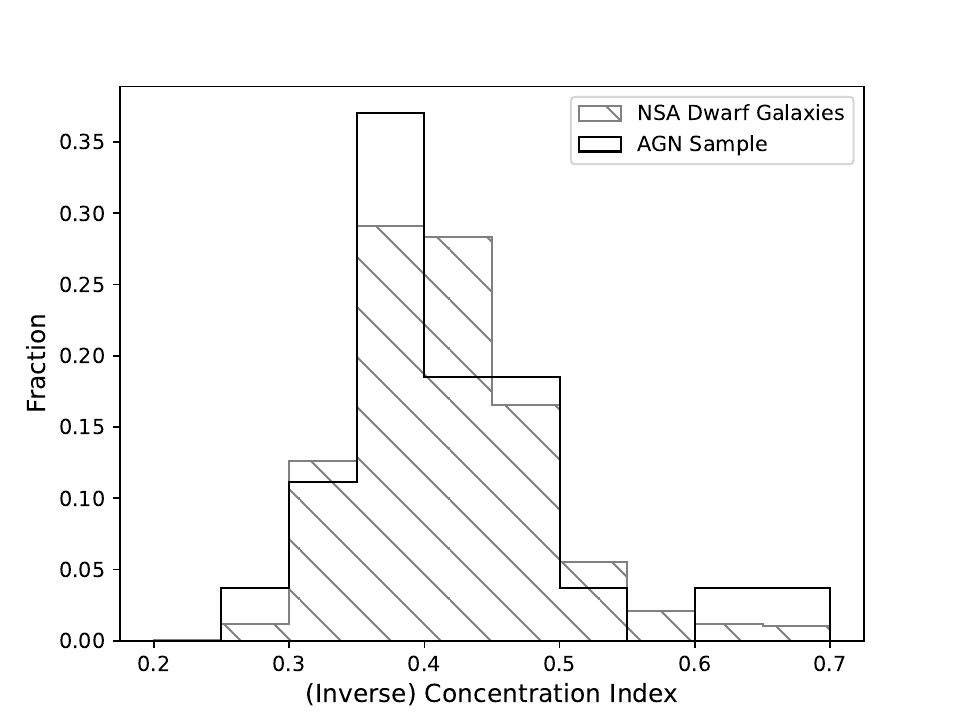}
 \caption{The (invese) concentration indices of the AGN sample compared to all dwarf galaxies in the NSA. The two populations are similar, with no clear difference in concentrations between AGN hosts and the parent dwarf galaxy population.}
\label{fig:conc_ind}
\end{figure}

We also examine the (inverse) concentration indices of the galaxies in our sample, defined as as $C=r_{50}/r_{90}$,  where $r_{50}$ and $r_{90}$ are the half-light and 90\%-light radii, respectively \citep{shimasaku2001}. Lower (inverse) concentration indices are indicative of more centrally concentrated galaxies. The X-ray AGN hosts mostly have (inverse) concentration indices that follow the general population of the dwarf galaxies, indicating that the AGN hosts are not systematically more compact than typical dwarf galaxies (Figure \ref{fig:conc_ind}).

Overall, the X-ray AGN sample shares many properties with the general dwarf galaxy populations, with only minor deviations, such as individual outliers in color and SFR. There are no clear correlations between the host galaxy properties and the presence of X-ray AGNs.


\subsection{X-ray Luminosities and Hardness Ratio}
\label{subsec:xray_lum}
\begin{figure}
\includegraphics[width=\columnwidth]{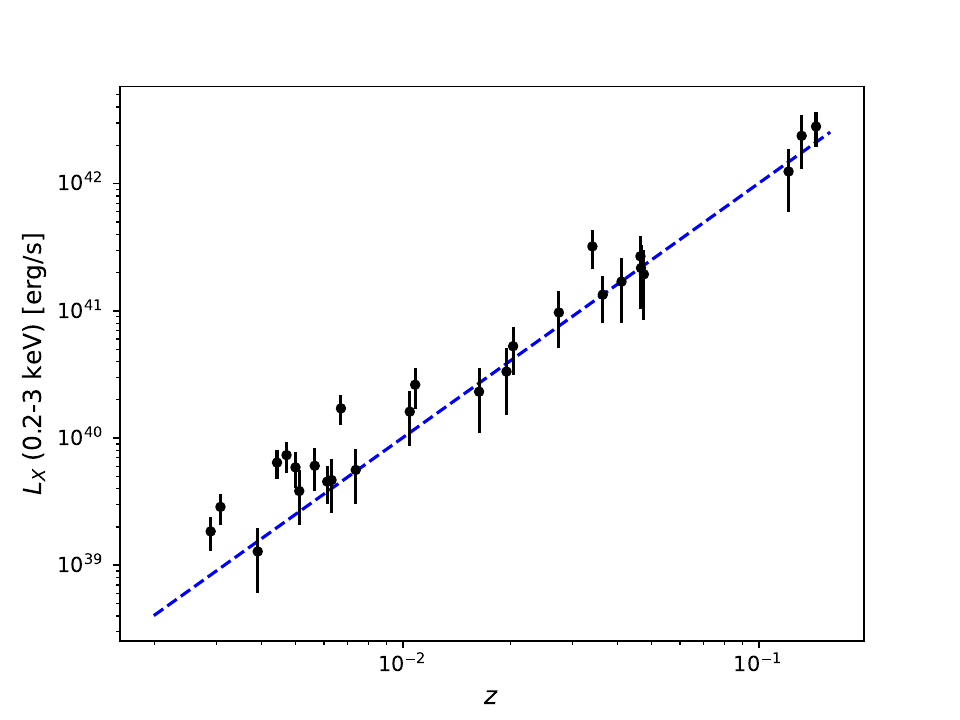}
 \caption{X-ray luminosity in the 0.2 - 2.3 keV band versus redshift for the sources in our AGN sample. The average detection limit of eRASS is shown as a dashed blue line.}
\label{fig:zlum}
\end{figure}

We analyze the X-ray luminosities of the sources in our final AGN sample as a function of redshift in Figure \ref{fig:zlum}. The observed positive correlation between luminosity and redshift is consistent with a flux-limited sample, since galaxies at smaller distances can be detected with lower luminosities compared to those at greater distances. Most of the galaxies in our sample have X-ray fluxes near 
the average flux limit of the eRASS survey \citep{merloni2024}. 

We further analyze the X-ray sources by studying their average hardness ratio. Given the low number of X-ray counts for many of the galaxies in our sample, hardness ratios for individual sources cannot be reliably calculated. Instead, we calculate the average hardness ratio of our sample, following the methodology of \cite{sacchi2024}. We define the average hardness ratio as S/M, where S is the total number of counts measured in the soft band ($0.5-1$ keV) and M is the total number of counts measured in the medium band ($1-2$ keV). {The average hardness ratio of our sample is found to be S/M $= 1.06^{+0.15}_{-0.11}$, which, assuming a power-law spectral model with a column density of $N_{\mathrm{H}} = 3 \times 10^{20} \; \mathrm{cm}^{-2}$, is consistent with a photon index of $\Gamma$ = $1.86^{+0.21}_{-0.17}$}
{This result is in agreement with AGN samples from eFEDS \citep{liu2022} and \cite{sacchi2024}, which report photon indices  of $\Gamma$ = 2 and $\Gamma$ = 1.8, respectively.}

\subsection{Positional Offsets}
\label{subsec:offsets}
\begin{figure}
\includegraphics[width=\columnwidth]{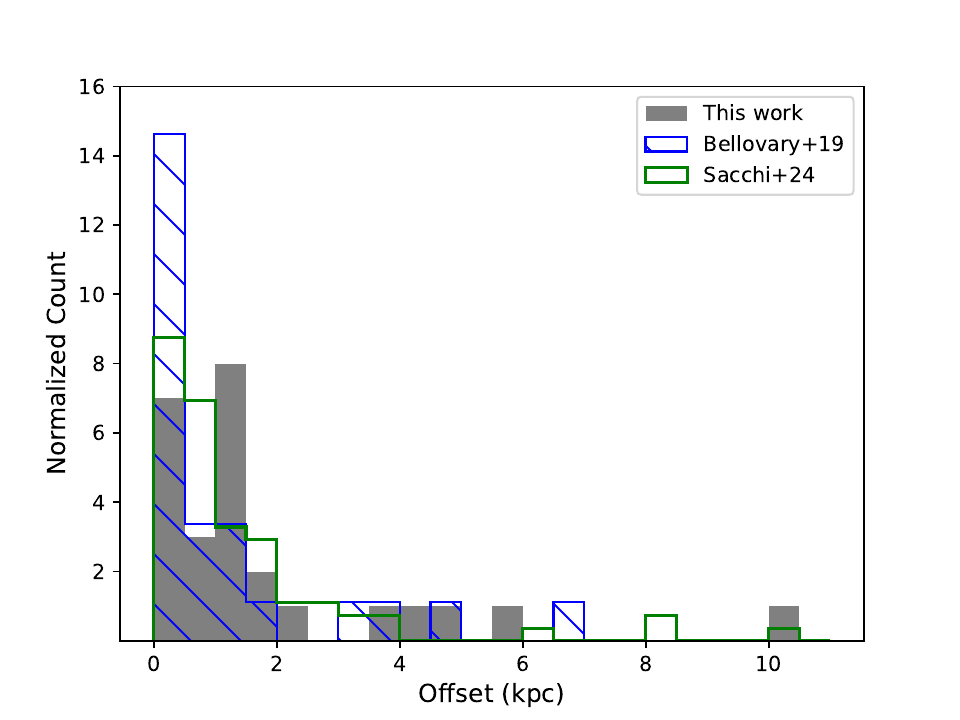}
 \caption{Histogram of the offsets of the X-ray sources from the optical center of their host galaxies. We compare our distribution of 27 AGNs in gray to the expected values from \cite{bellovary2019} in blue, and the X-ray-selected AGN sample from \cite{sacchi2024} in green. We find a larger number of sources with offsets between 1-2 kpc than expected, although we note that this histogram does not include the positional uncertainties of the X-ray source positions.
 }
\label{fig:offset_kpc}
\end{figure}

We analyze the positional offsets (in kpc) between the X-ray sources and the optical centers of the galaxies given by the NSA positions (Figure \ref{fig:offset_kpc}). The observed distribution of offsets is compared to the simulations from \cite{bellovary2019}, scaled to match our sample size. Overall, the offsets we find are in the range found in the simulations, although we find a larger fraction of sources within 1-2 kpc and a smaller fraction of sources within 0.5 kpc than in the \cite{bellovary2019} simulations.
{Our sample also differs from the AGN sample found in \cite{sacchi2024}, which had  a larger percentage of X-ray sources within 1 kpc of their host galaxies' optical centers than our sample (see Figure \ref{fig:offset_kpc}).}

We note, however, that the histogram does not include the positional uncertainties of the X-ray positions.
When we consider the uncertainties  of the X-ray positions, we find that 15 sources (56\%) are located within one times the X-ray positional uncertainty of the optical nucleus of their respective host galaxies, while 24 sources (89\%) fall within two times the positional uncertainty.

\begin{figure}
\includegraphics[width=\columnwidth]{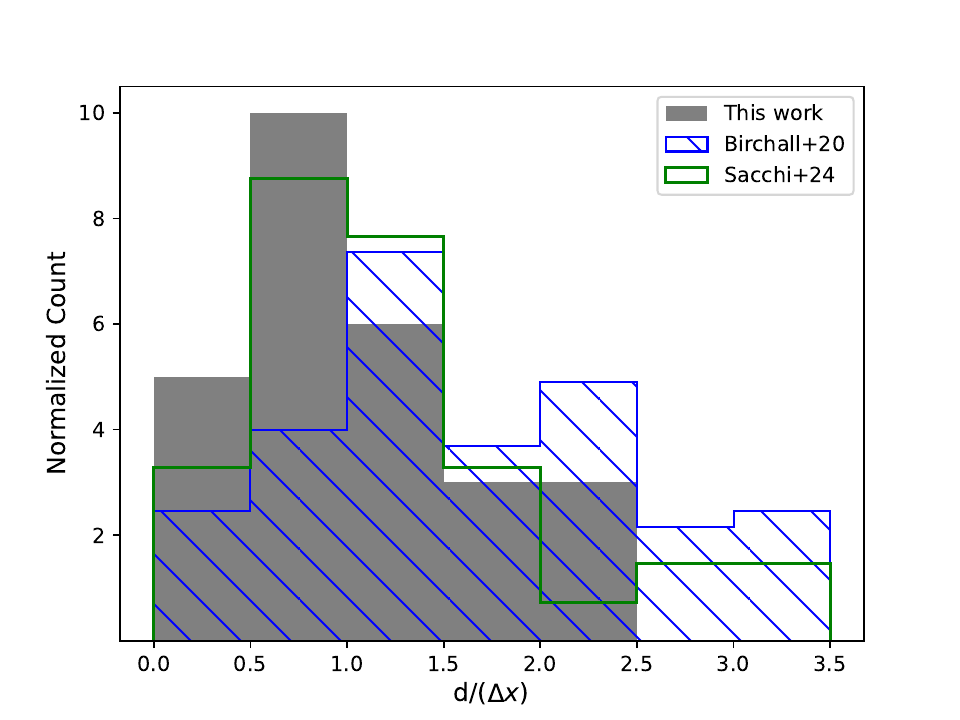}
 \caption{Histogram of the normalized offsets of the X-ray sources from the optical center of their host galaxies. We compare our AGN sample to the X-ray-seleccted AGN samples from \cite{birchall2020} and \cite{sacchi2024}.
 }
\label{fig:offset_x}
\end{figure}

{Figure \ref{fig:offset_x} presents the distribution of normalized offsets for the X-ray sources in our sample, where the normalized offset is defined as the separation between the X-ray source and the optical center, divided by the total positional uncertainty of the X-ray source. A comparison with the offset distributions of the X-ray AGNs in \cite{birchall2020} and \cite{sacchi2024} reveals that our sample exhibits a higher fraction of sources located within one times the X-ray positional uncertainty of the optical center. Furthermore, in contrast to these previous studies, our sample does not include any sources with offsets exceeding 2.5 times the X-ray positional uncertainty.}


Thus, the majority of the X-ray sources are consistent with nuclear origins, although a small fraction may represent wandering black holes. 
We note, however, that this classification is tentative, since the irregular morphology of dwarf galaxies can make it difficult to identify their centers \citep[e.g.,][]{pasetto2003,kimbrell2021,lazar2024}.

We find that several X-ray sources in our sample have multiple optical counterparts within the radii of their positional uncertainties. For example, while the positional uncertainty of the X-ray source of ID 16 significantly overlaps with its assigned host galaxy, the X-ray emission appears centered on a different red object nearby. 
While we have taken extensive measures to ensure accurate source associations, it remains uncertain in some cases which optical counterpart is physically linked to the X-ray emission. 
Higher-resolution X-ray observations or additional follow-up studies are required to definitively determine the true host galaxies of these X-ray sources.



\subsection{Black Hole Masses and Eddington Ratios}
\label{subsec:mass_edd}
We obtain rough estimates for the BH masses of the AGNs in our sample by applying the scaling relation  from \cite{reines2015}, calibrated
using local galaxies hosting broad-line AGNs, including dwarfs: 
\begin{equation}
\label{eq:bh_mass}
    \mathrm{log}\left(\frac{M_{BH}}{M_\odot}\right) = 7.45 + 1.05\; \mathrm{log}\left(\frac{M_{*}}{10^{11} M\odot}\right) \pm 0.55.
\end{equation}
Using this relation, we estimate BH masses in the range of  $\sim 10^{4} - 10^6 M_\odot$ for our sample. 

We estimate the Eddington ratios ($\lambda_{Edd}$)  by comparing the BH masses to hard X-ray luminosities ($2–10$ keV) 
We estimate hard X-ray luminosites from the observed $0.2-2.3$ keV luminosities using a correction factor of -0.06 dex, obtained from PIMMS, using the same assumptions as in \S\ref{subsec:HMXBs_XRBs}.
Following previous studies  (e.g., \citealt{aird2012}; \citealt{birchall2020}; \citealt{birchall2022}; \citealt{zou2023}), 
we compute $\lambda_{Edd}$ using the following equation:
\begin{equation}
    \lambda_{Edd} = \frac{K \times \mathrm{L_{2-10\; keV}}}{1.3 \times 10^{38} M_{BH}}
\end{equation}
where $K$ is the bolometric correction, which converts the hard 
X-ray luminosity into bolometric luminosity, and $M_{BH}$ is the BH mass calculated using Equation \ref{eq:bh_mass}. We adopt $K \approx 16.7$, consistent with \cite{bykov2024}, who derived this value from the work of \cite{duras2020}.

\begin{figure}
\includegraphics[width=\columnwidth]{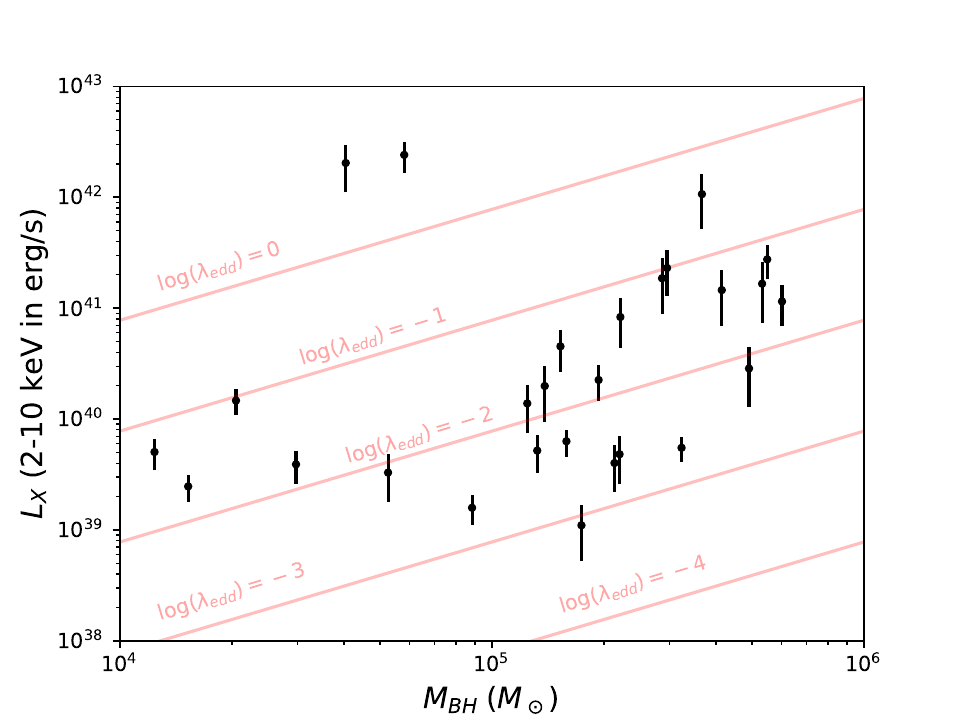}
 \caption{BH mass versus X-ray luminosity of the 27 galaxies in our AGN sample. Lines of constant Eddington ratio are shown with light red lines.  Two galaxies (IDs 7 and 18)  have Eddington ratios above $\lambda_{edd}$ = 1.}
\label{fig:edd_ratio}
\end{figure}

The majority of our sources have $\lambda_{Edd}$ values between $10^{-3}$ and $10^{-1}$ (see Figure \ref{fig:edd_ratio}). However, 2 galaxies (IDs 7 and 18)
have values that are super-Eddington, with $\lambda_{Edd} = 6.5$ and $\lambda_{Edd} = 5.3$, respectively. 
These values should be taken with caution, since bolometric corrections can vary significantly at high Eddington ratios \citep{duras2020}, and the spectral energy distribution may deviate from a simple power law. Additionally, the scaling relation between stellar mass and BH mass is highly uncertain in the low-mass regime (\citealt{suh2020}; \citealt{mezcua2023}).
Nevertheless, our findings suggest a range of accretion rates across the sample, highlighting the diversity of X-ray AGN activity in dwarf galaxies. 

\begin{deluxetable*}{ccccccccccccc}
\tablenum{1}
\tablecaption{AGNs in eRASS and the NSA}
\tablewidth{0pt}
\tablehead{
\colhead{ID} &
\colhead{Name} & 
\colhead{NSA ID} & 
\colhead{R.A.} &
\colhead{Dec.} & 
\colhead{R.A.$_{X}$} &
\colhead{Dec.$_{X}$} &
\colhead{$\sigma_{X}$} &
\colhead{Offset} &
\colhead{$z$} & 
\colhead{$r_{50}$}&
\colhead{log($M_{*}$)} &
\colhead{log(L$_{X}$)} 
}
\colnumbers
\startdata
        1 & J0851+3935 & 229068 & 132.8576 & 39.5949 & 132.8580 & 39.5949 & 7.36 & 1.10 & 0.0411 & 2.15 & 9.25 & 41.23(0.23) \\  
        2 & J0910+0712 & 648204 & 137.6744 & 7.2070 & 137.6738 & 7.2045 & 4.51 & 9.25 & 0.0056 & 7.81 & 8.78 & 39.78(0.16) \\  
        3 & J0923+2455 & 492135 & 140.9712 & 24.9280 & 140.9695 & 24.9287 & 6.63 & 6.10 & 0.0475 & 3.89 & 9.36 & 41.29(0.24)\\  
        4 & J0936+2734 & 411683 & 144.1240 & 27.5788 & 144.1219 & 27.5789 & 5.72 & 6.69 & 0.0274 & 4.34 & 9.00 & 40.99(0.21) \\  
        5 & J0940+0357 & 623354 & 145.1801 & 3.9589 & 145.1792 & 3.9598 & 6.69 & 4.47 & 0.0051 & 6.96 & 8.40 & 39.58(0.20) \\  
        6 & J0956-0304 & 623559 & 149.1794 & -3.0683 & 149.1784 & -3.0694 & 8.53 & 5.33 & 0.0464 & 4.37 & 9.11 & 41.43(0.19)\\  
        7 & J1001-0307 & 649900 & 150.2682 & -3.1234 & 150.2689 & -3.1226 & 5.28 & 3.91 & 0.1318 & 27.51 & 8.29 & 42.38(0.20)\\  
        8 & J1047+2208 & 511046 & 161.7725 & 22.1356 & 161.7718 & 22.1359 & 6.36 & 2.67 & 0.0195 & 2.24 & 9.32 & 40.52(0.24) \\  
        9 & J1049+3246 & 432210 & 162.3545 & 32.7726 & 162.3540 & 32.7749 & 5.56 & 8.44 & 0.0063 & 14.37 & 8.98 & 39.67(0.20)\\  
        10 & J1052+1947 & 656026 & 163.1330 & 19.7925 & 163.1322 & 19.7929 & 3.89 & 3.11 & 0.0047 & 12.78 & 8.86 & 39.87(0.12) \\  
        11 & J1057+3615 & 438664 & 164.4459 & 36.2608 & 164.4435 & 36.2605 & 3.35 & 6.89 & 0.0031 & 13.76 & 7.89 & 39.46(0.12) \\  
        12 & J1136+2526 & 519914 & 174.0616 & 25.4354 & 174.0614 & 25.4371 & 5.81 & 6.31 & 0.0104 & 7.81 & 8.76 & 40.21(0.20) \\  
        13 & J1136+1252 & 323229 & 174.2024 & 12.8778 & 174.2028 & 12.8761 & 4.61 & 6.22 & 0.0340 & 4.05 & 9.37 & 41.51(0.15)\\  
        14 & J1141+3225 & 662645 & 175.2812 & 32.4270 & 175.2812 & 32.4262 & 4.59 & 2.85 & 0.0067 & 5.23 & 8.01 & 40.23(0.11) \\  
        15 & J1157+3220 & 664492 & 179.3822 & 32.3417 & 179.3816 & 32.3399 & 3.98 & 6.68 & 0.0108 & 4.75 & 8.94 & 40.42(0.15)\\  
        16 & J1157+0158 & 627532 & 179.4727 & 1.9753 & 179.4720 & 1.9755 & 5.78 & 2.38 & 0.0466 & 4.42 & 9.10 & 41.34(0.23) \\  
        17 & J1203+0630 & 328898 & 180.7618 & 6.5012 & 180.7608 & 6.5007 & 6.46 & 4.34 & 0.0050 & 4.72 & 7.80 & 39.77(0.14) \\  
        18 & J1212+0113 & 580727 & 183.0501 & 1.2171 & 183.0499 & 1.2171 & 4.54 & 0.69 & 0.1446 & 0.87 & 8.44 & 42.45(0.13)\\  
        19 & J1216-0345 & 666554 & 184.0527 & -3.7605 & 184.0546 & -3.7601 & 5.71 & 7.00 & 0.1211 & 3.72 & 9.20 & 42.09(0.23)\\  
        20 & J1219+1713 & 538221 & 184.8695 & 17.2304 & 184.8706 & 17.2303 & 6.63 & 3.76 & 0.0039 & 14.82 & 8.90 & 39.11(0.23)\\  
        21 & J1221+1738 & 538328 & 185.2969 & 17.6387 & 185.2966 & 17.6374 & 4.52 & 4.68 & 0.0074 & 6.54 & 8.99 & 39.75(0.20)\\  
        22 & J1226+2826 & 480501 & 186.5367 & 28.4336 & 186.5365 & 28.4348 & 8.18 & 4.39 & 0.0164 & 2.39 & 8.80 & 40.36(0.23)0\\  
        23 & J1245+2707 & 481205 & 191.3219 & 27.1256 & 191.3238 & 27.1259 & 3.34 & 6.07 & 0.0044 & 9.51 & 9.15 & 39.81(0.11)\\  
        24 & J1248+0159 & 69774 & 192.2487 & 1.9883 & 192.2494 & 1.9883 & 4.57 & 2.67 & 0.0204 & 5.77 & 8.84 & 40.72(0.18)\\  
        25 & J1253+0427 & 150802 & 193.3107 & 4.4632 & 193.3113 & 4.4648 & 4.15 & 5.86 & 0.0029 & 7.80 & 8.62 & 39.27(0.13)\\  
        26 & J1351+1111 & 347965 & 207.8745 & 11.1858 & 207.8739 & 11.1856 & 4.58 & 2.26 & 0.0364 & 2.41 & 9.41 & 41.13(0.18)\\  
        27 & J1439+0244 & 74327 & 219.9926 & 2.7478 & 219.9952 & 2.7478 & 3.78 & 9.24 & 0.0061 & 11.76 & 8.16 & 39.66(0.14)\\  
\enddata
\tablecomments{Column 1: galaxy identification number assigned in this paper. Column 2: galaxy name. Column 3: NSA ID (version v1\textunderscore0\textunderscore1). Column 4: R.A. of galaxy from the NSA, in units of degrees. Column 5: dec. of galaxy from the NSA, in units of degrees.  Column 6: R.A. of galaxy from eRASS, in units of degrees. Column 7: dec. of galaxy from eRASS, in units of degrees. Column 8: positional uncertainty of the X-ray position from eRASS, in units of arcsec. Column 9: offset between the optical center from the NSA and the X-ray source from eRASS, in arcsec. Column 10: redshift given by NSA. Column 11: half-light radius from the NSA, in arcsec. Column 12: log galaxy mass given by the NSA, in units of $M_\odot$. Column 13: {log X-ray luminosity in the 0.2 - 2.3 keV band in eROSITA in units of erg s$^{-1}$, with errors in parentheses. Luminosities are calculated using fluxes from {eRASS1} and redshifts from the NSA.}
}
\end{deluxetable*}
\label{tab:erass_agns}
\subsection{Surveys at Other Wavelengths}
\label{subsec:wavelength_regimes}
In this section, we investigate whether any dwarf galaxies in our AGN sample exhibit additional evidence of AGN activity in available multi-wavelength surveys. We note that targeted observations were not conducted for these objects; therefore, the absence of detections should not be interpreted as evidence against the presence of AGNs. This is particularly relevant for relatively low-luminosity AGNs, which may fall below the detection limits of the surveys considered.

\subsubsection{Radio Observations}
We crossmatch the X-ray source positions with the positions of radio sources from  the Very Large Array Sky Survey (VLASS). Only source ID 13 has a radio counterpart within twice the X-ray positional uncertainty, although this source is still found beyond one times the X-ray positional uncertainty from the X-ray source.
This radio source was previously identified as a radio AGN by \cite{reines2020}
(ID 64 in their paper), 
but follow-up spectroscopy revealed it to be a background
quasar at z = 0.761 (Sturm et al., in preparation). 
We note, however, that since the radio source lies more than one times the X-ray positional uncertainty away from the X-ray source, the X-ray emission  and the radio emission could originate from uncorrelated sources.

We also compare our sample to the Faint Images of the Radio Sky at Twenty-Centimeters (FIRST, version 2013 Jun 5) catalog and identify 6 X-ray sources (IDs 2, 16, 13, 15, 23, and 25) with corresponding radio sources.
For these sources, we calculate the IR-radio correlation parameter $q$ to distinguish between radio AGNs and star formation:
\begin{equation}
q = \mathrm{log}\left(\frac{\mathrm{L_{TIR}}}{ 3.75\times10^{12}\; \mathrm{L_{1.4\;GHz}}}\right)    
\end{equation}
Here, $\mathrm{L_{TIR}}$ is the total infrared luminosity, which we derive using the W4 magnitude for the closest WISE source to the X-ray source.
We compare the $q$ values of our sources to the AGN threshold of $q < 1.94$ established by \citet{eberhard2025}, {who derived this criterion using a sample of NSA galaxies (including dwarfs) by identifying the $q$ value below which the radio emission became inconsistent with that expected from SF. Of the 7 galaxies in our final sample with FIRST detections, only ID 13
has a $q$ value inconsistent with SF}. However, as previously noted, its radio emission originates from a background quasar. The remaining sources do not meet the AGN criterion above, 
suggesting that while these galaxies may host X-ray AGNs, the detected radio emission from FIRST (on 5\arcsec\ scales) is dominated by star formation.



\subsubsection{Mid-IR Observations}
\begin{figure}
\includegraphics[width=\columnwidth]{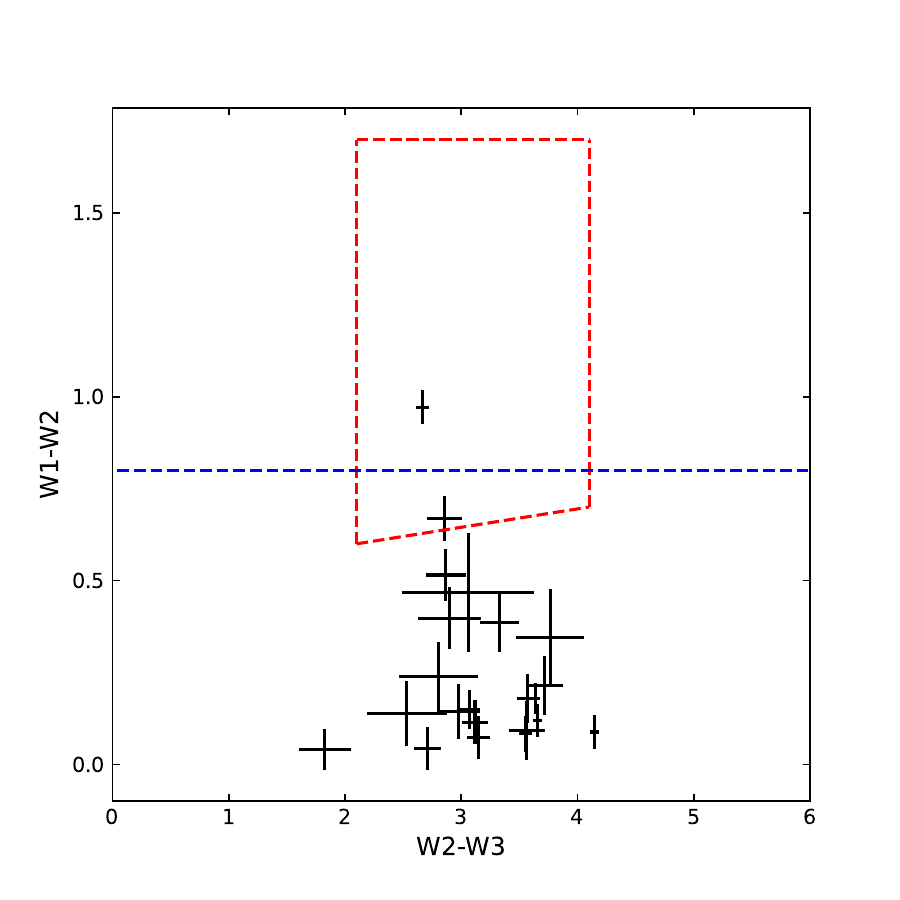}
 \caption{WISE colors of the galaxies in our final AGN sample with reliable WISE detections. The \cite{stern2012} cutoff is shown as in blue and the \cite{jarrett2011} selection box is shown in red. We find that 2 of the galaxies (IDs 6 and 7) have colors consistent with mid-IR AGNs.}
\label{fig:wise_color}
\end{figure}

We consider the WISE detections of the sources in our sample to see if they have IR colors consistent with mid-IR AGNs. We remove galaxies without W3 detections and those with poor W3 detections (S/N $<$ 2), and plot the remaining galaxies in a WISE color-color plot (Figure \ref{fig:wise_color}). Two galaxies in our sample (IDs 6 and 7)
exhibit mid-IR colors consistent with AGNs, since they are within the \cite{jarrett2011} selection box and/or above the \cite{stern2012} cutoff. 

\subsubsection{Optical Observations}
\begin{figure}
\includegraphics[width=\columnwidth]{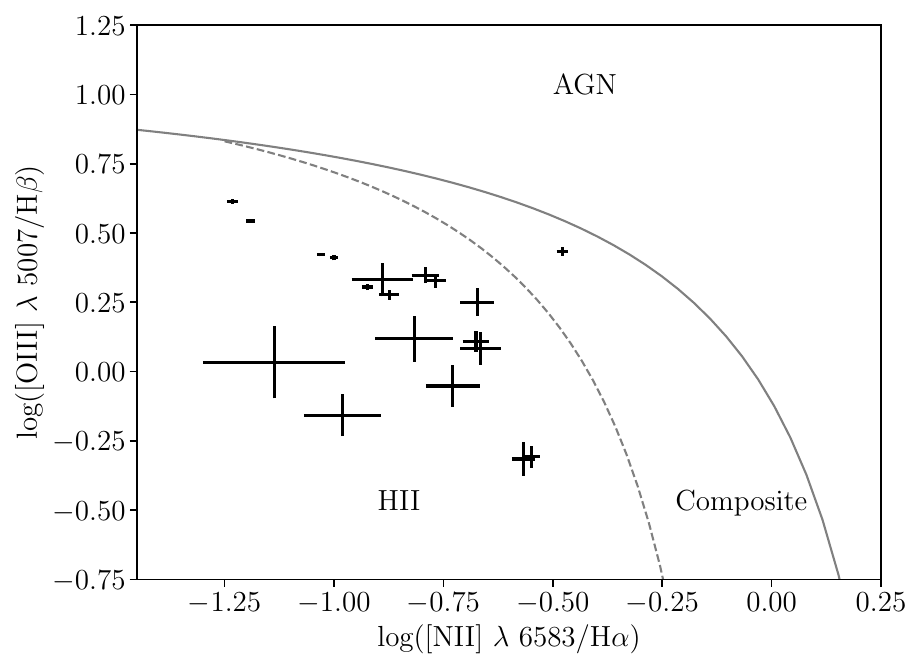}
 \caption{Location of the 22 galaxies in our AGN sample with SDSS spectra on the BPT diagram. Only one galaxy (ID 1) is found outside the HII region of the diagram. The dividing lines in the diagram come from the classification scheme outlined in \cite{kewley2006}.}
\label{fig:bpt}
\end{figure}

Of the 27 dwarf galaxies in our AGN sample, 22 have spectra that have been fitted in the SDSS. We compare the galaxy emission lines in the [OIII]/H$\beta$ versus [NII]/H$\alpha$ 
BPT diagram \citep{baldwin1981} in Figure \ref{fig:bpt}. Only one galaxy (ID 1) has emission lines consistent with a composite galaxy. The remaining galaxies have optical lines consistent with SF. This further shows the ability of X-ray emission at finding AGNs located in galaxies with considerable star formation. 
However, we note that the SDSS optical fibers do not always overlap with the X-ray positions (see Figure \ref{fig:decals_pics}). As such, for these galaxies, the optical classification is not expected to necessarily match the classification from the X-ray emission.

\subsection{Previously Identified AGNs}
\label{subsecc:previous_agns}

\subsubsection{Comparison to \cite{sacchi2024}}
\label{subsubsec:sacchi}
We compare our findings with those from \cite{sacchi2024}, who also used the {eRASS1} catalog to identify X-ray bright AGNs in dwarf galaxies in the western hemisphere. 
Eight of the 27 galaxies in our final AGN sample (IDs 2, 5, 9, 10, 15, 20, 21, and 22)
were also identified by \cite{sacchi2024}.
The remaining 19 galaxies were not included in their study, primarily due to the differences between the NSA and the HECATE catalogs that were used to define the parent sample of dwarf galaxies of this paper and \cite{sacchi2024}, respectively. Specifically,
\begin{itemize}
    \item Six galaxies (IDs 3, 6, 7, 16, 18, and 19) have redshifts in the NSA exceeding the redshift limit of the HECATE catalog ($z<0.047$). 
    \item  Nine galaxies (IDs 4, 11, 12, 13, 14, 17, 24, 26, and 27) lack mass estimates in HECATE. 
    \item One galaxy (ID 8) has a stellar mass in the HECATE catalog greater than the dwarf galaxy mass cutoff ($3 \times 10^{9} M_\odot$).
\end{itemize}
The discrepancies in stellar masses between the catalogs arise from the differences in the photometric methods used to calculate mass, with 
NSA masses being derived from UV and optical data and HECATE masses being derived from near-IR data. 

{The 3 remaining galaxies (IDs 1, 23, and 25) were in the HECATE catalog but failed to be classified as AGNs by \cite{sacchi2024}. 
IDs 1 and 23 
are included in MILLIQUAS catalog, but were excluded by \cite{sacchi2024} for being quasars. Upon inspection of the MILLIQUAS catalog, these sources are classified as AGNs, but not as quasars, and have accurate redshifts. For this reason, we keep them in our sample, though we note that they have previously been identified as AGNs.
ID 25 was removed from \cite{sacchi2024} for having an X-ray luminosity consistent with an HMXB; it had a smaller distance in HECATE than in the NSA, resulting in a smaller X-ray luminosity consistent with an individual HMXB (i.e., $\mathrm{L_{X}}<10^{39}$ erg s$^{-1}$).}

{Our results underscore the complementary nature of this study to that of \cite{sacchi2024}, and highlight how differences in parent galaxy catalogs, redshift limits, and stellar mass estimation methods can significantly impact AGN identification in dwarf galaxies. While there is partial overlap between the samples, the majority of our identified AGNs were excluded from \citet{sacchi2024} due to catalog-specific selection effects rather than intrinsic differences in the sources themselves. This emphasizes the value of employing diverse datasets and selection strategies to build a census of AGNs in low-mass galaxies.}

\subsubsection{Comparison to the Remaining Literature}
\label{subsubsec:remaining_lit}
Two of the 10 galaxies identified in \cite{sacchi2024} and this paper (IDs 5 and 21) 
were also studied in other papers.
ID 5 was identified as an X-ray bright AGN by \cite{latimer2021} (ID 2 in their paper) 
and was confirmed as being an X-ray AGN by \cite{sanchez2024}. 
ID 21 was identified {as a BH in \cite{lemons2015}} and was further studied by \cite{thygesen2023}, where the source was determined to likely not be an AGN because of its offset from the galactic nucleus. However, since \cite{bellovary2019} predicts that a large fraction of AGNs in dwarf galaxies could be offset, it could still be an AGN, as noted by \cite{sacchi2024}. 

{Four} additional galaxies were previously identified in the literature as being AGN candidates: 

\begin{itemize}
    \item ID 1 was classified as as a composite galaxy on the BPT diagram in \cite{reines2013} 
    {(ID 48 in their paper)} and was confirmed to be an AGN with X-ray  emission by \cite{baldassare2017} 
    \item ID 13 was detected in high-resolution (0.25\arcsec) VLA observations \citep[][ID 64 in their paper]{reines2020}. ID  13 was identified as a radio AGN, though its radio emission was found to originate from a background quasar.
    \item {ID 14 was studied using Chandra, VLA and HST observations in \cite{wang2025}, and was found to be a likely candidate for a massive BH.}
    \item ID 23 was  identified as a plausible BH in \cite{latimer2019} {(Haro 9 in their paper)} with a mass of $M_{BH} \sim 10^{4.8\pm1.1}$.
\end{itemize}

In total, {12} of the galaxies in our AGN sample 
were previously identified as candidate AGN hosts. The remaining {15} represent novel AGN candidates identified in this study.

\section{Conclusions} 
\label{sec:Conclusions}
We conducted a crossmatching analysis between X-ray sources from the western half of the eRASS1 catalog and dwarf galaxies from the NSA, identifying 68 X-ray source-dwarf galaxy pairs. We removed likely background sources and galaxies with dubious stellar masses. Additionally, we removed 8 likely XRBs using 3 different XRB {luminosity scaling relations} from the literature. After the removal of these sources, our final sample consists of 27 X-ray-selected AGN candidates in dwarf galaxies, 15 of which are not previously reported in the literature. {An analysis of the parent sample revealed that $\sim 20$ of the 68 crossmatched sources could plausibly be explained by ULXs, 8 of which were likely removed with the XRB cuts. While up to $\sim 12$ ULXs could remain in our final sample, the 9 sources with $L_X > 10^{41} \; \mathrm{erg \; s^{-1}}$ are unlikely to be ULXs.} 

Our analysis of the AGN candidate host galaxies revealed that their stellar masses, SFRs, and concentration indices are broadly consistent with those of the parent dwarf galaxy population. No clear correlations were observed between host galaxy properties and the likelihood of hosting an X-ray-detected AGN. Additionally, {estimating BH masses from the scaling with stellar mass
given by \cite{reines2015}}, the majority of X-ray AGN candidates in our sample have Eddington ratios in the range $\sim 10^{-3}<\lambda_{Edd} <10^{-1}$, with a few sources accreting near or above their Eddington limit. These findings reinforce the effectiveness of X-ray observations in identifying AGNs across a wide range of Eddington ratios and host galaxy properties, including systems with significant ongoing star formation. 

We examined whether the galaxies were identified as AGN hosts across multiple wavelength regimes using various survey data. Seven galaxies were detected in the radio regime; however, only one exhibits radio emission {inconsistent with SF}, and this source has been confirmed as a background quasar. Three galaxies displayed mid-IR colors indicative of AGN activity. Optical spectroscopic classification using the BPT diagram indicated that all but one of the galaxies have emission line ratios consistent with star-forming regions, with the remaining source classified as a composite galaxy. While the majority of these galaxies do not exhibit strong evidence of AGN activity based on radio, mid-IR, and optical diagnostics, X-ray classifications are not always consistent with optical or IR criteria. Notably, X-ray observations have been shown to  consistently find AGNs that are classified as star-forming by optical AGN diagnostics \citep[e.g.,][]{birchall2020}. Thus, the absence of AGN signatures at other wavelengths does not necessarily preclude the presence of an AGN. Further multiwavelength analysis, particularly with higher-resolution observations, will be essential for accurately constraining the nature of these X-ray sources.



The AGN {candidate} sample presented here is a valuable resource for follow-up observations, offering new opportunities to probe the properties of BHs in dwarf galaxies. This work contributes to our understanding of BH demographics in low-mass systems and lays the foundation for future studies of BH growth and feedback in the low-mass regime.

\section*{Acknowledgments}
We thank the anonymous reviewer for their constructive comments which helped improve and clarify the manuscript.
A.E.R. gratefully acknowledges support for this work provided by NSF through CAREER award 2235277.

This work is based on data from the eROSITA telescope aboard the Spectrum-Roentgen-Gamma (SRG) observatory. eROSITA was developed under the leadership of the Max Planck Institute for Extraterrestrial Physics (MPE), with contributions from the German Aerospace Center (DLR), the Institute for Astronomy and Astrophysics of the University of Tübingen (IAAT), and other German institutions. The SRG observatory was developed by Roscosmos in collaboration with the Russian Academy of Sciences’ Space Research Institute (IKI). The eROSITA data used in this work were obtained from eRASS DR1. The authors acknowledge the eROSITA consortium for its support and efforts in mission operations, calibration, and data processing.

\section*{Appendix}
We present images (Figure A1) and properties (Table A1) of the galaxies that were removed from our sample for having X-ray emission consistent with XRB populations. 
The galaxies are included here to provide the scientific community with potential AGN candidates, albeit with lower confidence than those in our final AGN sample. Notably, IDs X1 and X2 were classified as AGNs by \citet{sacchi2024}, who applied XRB contamination criteria based solely on the prescriptions of \citet{lehmer2010}.

\renewcommand{\thefigure}{A1}
\begin{figure*}
\includegraphics[width=\textwidth]{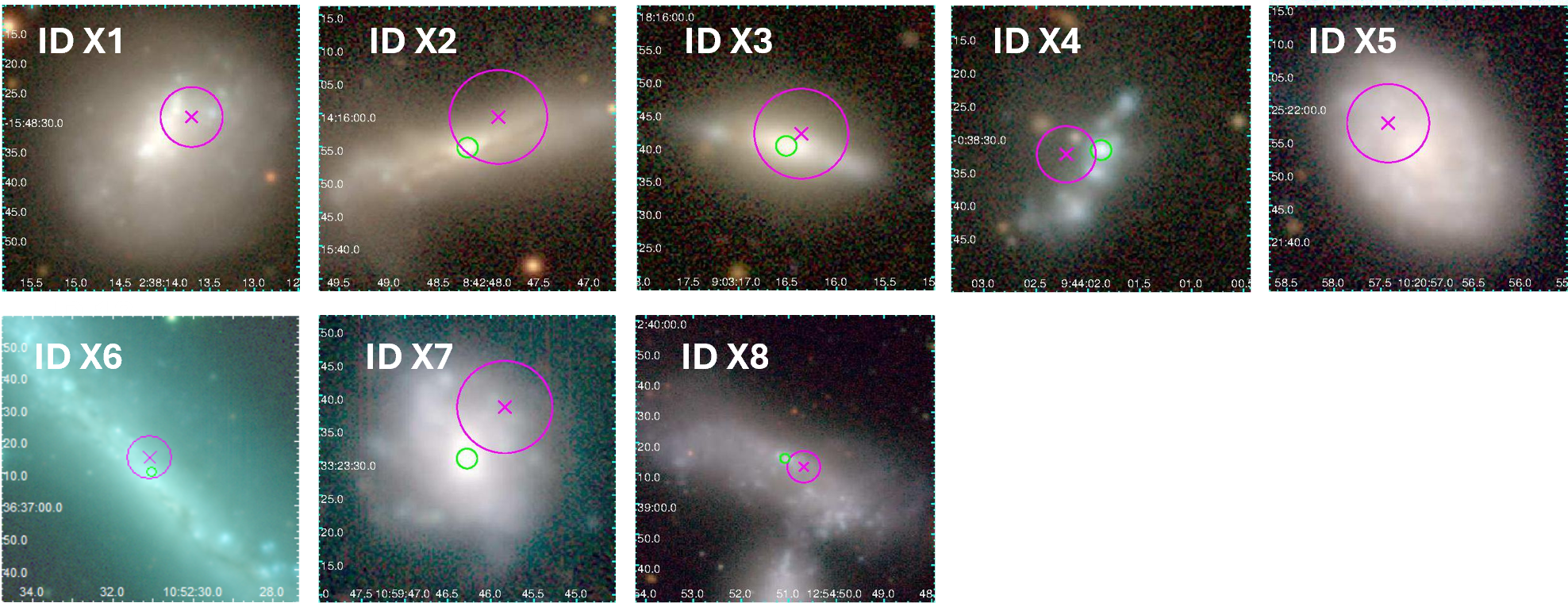} 
\label{fig:decals_xrbs}
\parbox{\textwidth}{\caption{DECaLS $grz$ images of the 8 galaxies in our sample with X-ray emission consistent with XRBs. Magenta crosses and circles show the positions and positional uncertainties of the X-ray sources from {eRASS1}. Green circles indicate the locations of the SDSS spectral fibers (with 1.5\arcsec radii) for the galaxies with SDSS spectra. }}
\end{figure*}

\begin{deluxetable*}{ccccccccccccc}
\tablenum{A1}
\tablecaption{Galaxies in eRASS and the NSA with XRB-consistent Emission}
\tablewidth{0pt}
\tablehead{
\colhead{ID} &
\colhead{Name} & 
\colhead{NSA ID} & 
\colhead{R.A.} &
\colhead{Dec.} & 
\colhead{R.A.$_{X}$} &
\colhead{Dec.$_{X}$} &
\colhead{$\sigma_{X}$} &
\colhead{Offset} &
\colhead{$z$} & 
\colhead{$r_{50}$}&
\colhead{log($M_{*}$)} &
\colhead{L$_{X}$}  
}
\colnumbers
\startdata
        X1 & J023814.00-154833.0 & 619849 & 39.5590 & -15.8097 & 39.5571 & -15.8082 & 5.05 & 8.60 & 0.0047 & 6.73 & 8.82 & 0.23(0.07) \\   
        X2 & J084247.89+141552.9 & 647793 & 130.7009 & 14.2652 & 130.6996 & 14.2665 & 7.11 & 6.45 & 0.0079 & 11.85 & 9.32 & 0.44(0.26) \\   
        X3 & J090316.49+181539.8 & 489964 & 135.8188 & 18.2611 & 135.8182 & 18.2616 & 6.82 & 2.86 & 0.0121 & 4.03 & 9.38 & 1.13(0.61) \\   
        X4 & J094401.87-003832.1 & 50 & 146.0078 & -0.6423 & 146.0092 & -0.6424 & 4.26 & 5.05 & 0.0055 & 7.08 & 7.75 & 0.28(0.13) \\   
        X5 & J102057.12+252153.8 & 496587 & 155.2380 & 25.3650 & 155.2393 & 25.3660 & 5.94 & 5.47 & 0.0053 & 6.94 & 9.20 & 0.22(0.11) \\   
        X6 & J105229.20+363649.8 & 438511 & 163.1293 & 36.6195 & 163.1295 & 36.6207 & 6.78 & 4.59 & 0.0031 & 44.17 & 9.37 & 0.10(0.05) \\   
        X7 & J105946.27+332330.5 & 419941 & 164.9428 & 33.3918 & 164.9410 & 33.3940 & 6.94 & 9.50 & 0.0069 & 6.39 & 9.13 & 0.39(0.19) \\   
        X8 & J125450.90+023912.0 & 671634 & 193.7128 & 2.6541 & 193.7112 & 2.6534 & 5.17 & 6.52 & 0.0038 & 20.68 & 8.58 & 0.11(0.05) \\       
\enddata
\tablecomments{Same columns and units as Table 1. }
\end{deluxetable*}
\label{tab:erass_xrbs}

\bibliography{bib}{}
\bibliographystyle{aasjournal}

\end{document}